\documentclass[lettersize,journal]{IEEEtran}
\usepackage{amsmath,amsfonts}
\usepackage{algorithmic}
\usepackage{algorithm}
\usepackage{array}
\usepackage[caption=false,font=normalsize,labelfont=sf,textfont=sf]{subfig}
\usepackage{textcomp}
\usepackage{stfloats}
\usepackage{url}
\usepackage{verbatim}
\usepackage{graphicx}
\usepackage{cite,url,hyperref,lineno,microtype,amsmath,graphicx,graphics,hyperref}
\begin{document}

\title{Empirical Interpretation of the Relationship \\ Between Speech Acoustic Context and \\ Emotion Recognition}

\author{Anna Ollerenshaw, Md. Asif Jalal, Rosanna Milner, Thomas Hain \\
Speech and Hearing Research Group, Department of Computer Science, University of Sheffield, UK
\thanks{This work was partly supported by VoiceBase Inc. at the VoiceBase Research Center which is supported by Liveperson, Inc.}
}

\markboth{}%
{}

\IEEEpubid{}

\maketitle

\begin{abstract}
Speech emotion recognition (SER) is vital for obtaining emotional intelligence and understanding the contextual meaning of speech. Variations of consonant-vowel (CV) phonemic boundaries can enrich acoustic context with linguistic cues, which impacts SER. In practice, speech emotions are treated as single labels over an acoustic segment for a given time duration. However, phone boundaries within speech are not discrete events, therefore the perceived emotion state should also be distributed over potentially continuous time-windows. 

This research explores the implication of acoustic context and phone boundaries on local markers for SER using an attention-based approach. The benefits of using a distributed approach to speech emotion understanding are supported by the results of cross-corpora analysis experiments. Experiments where phones and words are mapped to the attention vectors along with the fundamental frequency to observe the overlapping distributions and thereby the relationship between acoustic context and emotion. This work aims to bridge psycholinguistic theory research with computational modelling for SER. 
\end{abstract}


\section{Introduction}\label{sec:intro}
\IEEEPARstart{S}{peech} emotion understanding and recognition (SER) is a complex research area, with modeling approaches that aim to adapt to speech variability, while reducing redundancy in acoustic and linguistic perceptual cue recognition. These approaches are particularly challenging to develop because the target labels or the perceived emotion states can be considered very subjective and biased by cultural and linguistic perception differences. Speech emotion, within the domain of SER, is typically represented by two approaches: categorical and dimensional. Speech acoustic segments can be treated as a categorical entity consisting of discrete emotions such as \textit{happy}, \textit{sad}, \textit{fear}, etc \cite{ekman1992argument}. In the categorical approach, annotators label audio segments as emotion categories and use them to model speech emotion. The dimensional approach proposes two fundamental dimensions, valence and arousal, to represent emotion at a given time \cite{russell1980circumplex}.

\begin{figure}[t!]
		\centering
		\includegraphics[width=0.75\columnwidth,clip]{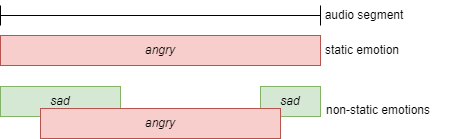}
		\caption{An example of distributed emotions where labelling an utterance as a single discrete category could be overlooking other perceived emotions}
		\label{fig:emo_pic}
\end{figure}

Typically, when a speech emotion corpora is created, each audio segment is labelled as a specific emotion category by the annotators, and it is assumed that the whole audio segment signifies that single emotion label \cite{enterface, iemocap, mosei}. It is theorised that the perceptual cues for phone boundaries and acoustic context are ambiguous as they share information for various emotion states \cite{doi:10.1080/02699931.2012.732559,ilie2011experiential}. The acoustic stimuli change in speech segments are distributed events and can therefore overlap.
From a psycholinguistic perspective, these distributed, continuous stimuli transitions constitute theories of human perception of SER \cite{doi:10.1080/02699931.2012.732559,ilie2011experiential}. The context cues can be of different lengths, and the perceptual acoustic context can be modelled with different length acoustic cues. Work from \cite{jalal2020empirical} shows that speech emotion can be modelled with small acoustic cues (200 ms). Therefore, the assumption that each acoustic speech segment is attributed to only one emotion state likely negatively impacts recognition performance. Multiple sub-emotions can be present depending on the contextual variation between different segment regions, as shown in Figure \ref{fig:emo_pic}. This paper focuses on acoustic perceptual cues and the implication of the length and the distribution of these cues over speech audio segments for SER.

Previous research for SER mainly focuses on modelling generalised emotion with different neural network architectures while adapting to speech variability and reducing redundancy for speaker invariance to improve SER capability \cite{Schuller2003, Zhang2016, mirsamadi2017automatic, huang2016attention, huang2014speech}. As the focus of current research has shifted towards embedding modelling and left-right context cues, work by \cite{kim2017deep} proposed a spatial representation learning method with CNNs, to model mid to long-term sequence dependencies. After the advent of the transformer architecture, the SER models focused more on transformer-based and multi-head fusion-based modelling approaches \cite{9054073, 9316758}. 

There remains a gap between the psycholinguistic and cognitive theories regarding speech emotion perceptual cues and the currently developed computational modelling methods. Research focusing on interpretability is still underdeveloped for SER models, particularly where the model's internal intricacies and representations with the corresponding acoustic segments can be explained. This work attempts to find a mutual accord with the theories of speech emotion perception cues across multiple disciplines and bridge the gap to speech emotion models. By projecting model attention weights across different time frames (based on various acoustic cues) of the acoustic segment, the emotion classification is observed to shift. Several corpora have been considered to demonstrate the task across various types of speech emotion data (acted, natural, elicited).

This manuscript is organised as follows. Section \ref{sec:intro} introduces the research domain and the particular research goal of the work, along with some background information. Section \ref{sec:context} discusses context modelling and introduces the idea of overlapping context regions and phone units. Section \ref{sec:linguist} presents the consonant-vowel (CV) boundaries and phonemic overlapped regions and their significance in speech emotion perception cues and recognition. Section \ref{sec:dnn} explains the underlying SER model for the interpretation framework and attention. Section \ref{sec::exp} describes the cross-corpus data, features, experimental framework, and presents the results and graphs. Section \ref{sec:disscussion} discusses the interpretation of the presented results and suggested directions for the development of future work.

\section{Context Modelling}\label{sec:context}

Context cues for speech emotion can be described as linguistic and paralinguistic. The linguistic aspects consist of semantic structure of the speech segment and the textual meaning. The nonverbal or paralinguistic aspects provide a rich source of perceptual context cues that facilitates projecting expressiveness in social discourse in both intra-cultural and cross-cultural scenarios \cite{wilson2006relevance,hoemann2019context}. Although verbal comprehension mainly dictates social discourse, perceptual context cues can deliver meaning and emotion independent of the verbal comprehension using the acoustic changes that influence the speech delivery \cite{scherer2001emotion, ThompsonBalkwill+2006+407+424}. Work in \cite{doi:10.1080/02699931.2012.732559} used psychoacoustic features (such as tempo, prosodic contour, loudness etc.) for modelling emotion and concluded that different emotional states have different perceptual cues and that they are subjective to individual contexts despite having a universal representation of emotion states. Furthermore, the acoustic contexts are not orthogonal, and the shared information/dimensions represents the redundant acoustic stimuli which provide context \cite{doi:10.1080/02699931.2012.732559,ilie2011experiential}. Naturally, if the acoustic stimuli changes, the perceptual context cue will also change accordingly. If the acoustic stimuli are redundant for the cues that define emotion states, these stimuli share overlapping regions. Typically, a `phone' is regarded as one of the smallest units of an acoustic speech sound. To explore the implication of the various stimuli regions, the phone boundaries should be explored. The CV boundaries for context cues are discussed in Section \ref{sec:linguist}.

The authors in \cite{stilp2020acoustic} have presented left context (referred to as ``forward effects'' by the authors), right context (referred to as ``backward effects'' by the authors), proximal context and distal acoustic context cues as a in acoustic events over time. The sensory attention emphasises the change among these acoustic stimuli, which maximises the potential information for facilitating speech perception \cite{kluender2019long}. The stimuli changes at a particular time over left-right time frames to reflect the emotion state and speech perception cue at that given point of time. Therefore, it can be assumed that emotion is a distributed event in acoustic segments, not a single discrete emotion category. To investigate this hypothesis, a simple computational model of left-right modelling with attention has been applied in Section \ref{sec:dnn}.


\section{Linguistic Boundaries}\label{sec:linguist}

Contextual cues, consisting of phonetic aspects for speech, can be used to aid the determination of the emotional state at a given time.
The phonological forms can have similarities and dissimilarities among the phone boundaries. A clear distinction has been found between the clusters of vowel and consonant phone datapoints by work from \cite{van2018role, fogerty2009perceptual}. The consonant phones play a decisive role in word meaning comprehension, such that removing initial prosodic variations in vowel phones (acoustic reduction) enhances word intelligibility \cite{van2018role}. However, contrasting studies showed that replacing intermittent consonants with noises or change in emphasis on vowels, increases the perceived intelligibility of words and sentences to human listeners \cite{owren2006relative, cole1996contribution}. It is argued that vowel phones are more responsible for defining the emotional state of the speech acoustics, and intelligibility due to stressed vowel regions and wide harmonic variations \cite{fogerty2009perceptual, Ladefoged195798,  doi:10.1121/1.5122190}. 

Furthermore, the harmonic variations and variations in the pitch within vowels, change the CV boundaries over time and contextual cues related to acoustic perception. These continuous perceptual context cues are distributed over CV boundaries in acoustic segments \cite{Ladefoged195798}. Thus it may be possible that at different left-right time-frames, different regions from the same acoustic segment may be categorised differently. This can be described as the relationship between perceptual CV cues with the acoustics, which has been referred to as acoustic-phonetic context for speech perception \cite{cooper1952some, miller1994internal}. The aim of this work is to understand the distributed nature of these perceptual acoustic cues which form intra-linguistic determinism between acoustic structure and meaning that humans perceive as emotion. Here, meaning and intelligibility are explored only from acoustic segments as no language model or external multi-modal data has been used.


\section{Model architecture: \textit{BLSTMATT}} 
\label{sec:dnn}

\label{ssec:blstmatt}

\begin{figure}[ht!]
 		\centering
 		\includegraphics[width=4cm]{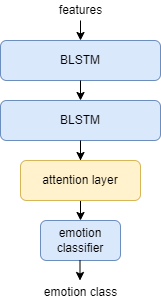}
 		\caption{The \textit{BLSTMATT} model pipeline consists of 2 bidirectional-LSTM layers, with an attention layer and linear classifier}
 		\label{fig:blstmatt}
 \end{figure}
 
The focus of this work is to explore perceptual acoustic cues and their relationship to current speech emotion recognition modelling. Developing and training large-scale SER models is out of the scope of this work as this approach is to determine the concept of this relationship. As the previously discussed related theories, regarding speech emotion perception, take into account past and future context, this can be modelled as a form of left and right acoustic cues. The chosen modelling approach utilises a bidirectional long short term memory (LSTM) neural network with a subsequent attention layer, referred to as \textit{BLSTMATT}. An overview of the model structure is displayed in Figure \ref{fig:blstmatt}.
 
LSTM networks are unable to exploit the future context and instead they solely focus on the temporal order of the sequence, whereas bidirectional LSTMs \cite{Graves2013} comprise of an additional layer of hidden connections which allows temporal information to pass in the opposite direction in order to exploit future and past contextual information \cite{Schuster1997}. 
The hidden connections $\mathbf{h}^n$ are iteratively compiled:
\begin{equation}
    h^n_t=\mathcal{H}(\mathcal{W}_{h^{n-1}h^n} h_t^{n-1}+\mathcal{W}_{h^nh^n}h^n_{t-1} +b_h^n)
\end{equation}
where $\mathcal{W}$ defines the weight matrices, $\mathcal{H}$ represents the hidden layer function and $b$ refers to the bias vector. 
Using this approach, temporal feature distribution over the sequence can be obtained, which is more effective for SER tasks \cite{mani2021stochastic}.

The attention mechanism enables computation of longer-term inter-sequence dependencies. The additive method for computing attention from \cite{beard2018multi} is applied for this approach. Utilising the global mean, the attention mechanism enables the network to attend to specific parts of itself which in turn captures global information. The non-linearity $tanh$ is used to multiply the global mean over the whole temporal vector which computes the positional dependency of each element. $\mathcal{H}$ denotes the matrix of output vectors from the LSTM layer, by summing the average time of $\mathcal{H}$ across contextual modalities, the shared memory matrix $\mathcal{M}$ can be formed by repetition until it matches the dimension of $\mathcal{H}_s$, where $\mathbf{s}$ refers to the view for the context. Where $T$
refers to iterations, $\mathcal{V}$ denotes the parameters controlling the influence within the view and from the shared memory:
\begin{equation}\label{eq:gamma}
    \gamma^{(\tau)} = tanh(\mathcal{V}^{(\tau)}_{s1} tanh(\mathcal{H}_{s1}))\cdot tanh(\mathcal{V}^{(\tau)}_{s2}\mathcal{M}^{(\tau}))
\end{equation}
\begin{equation}\label{eq:alpha}
    \alpha^{(\tau)}_s = \mathcal{V}^{(\tau) T}_{s3} \gamma^{(\tau)} 
\end{equation}
$\mathcal{V}_{s1}$, $\mathcal{V}_{s2}$ and $\mathcal{V}_{s3}$ are parameters used to compute the attention strength $\alpha$. 
%

The \textit{BLSTMATT} model setup consists of 2 x 512 dimension hidden layers feeding into an attention layer, which computes a 128 dimension context vector. For classification, the network uses a fully-connected linear layer which projects the attention output. In order to classify over the number of emotions in the target, the output is normalised with a $softmax$ layer before the loss is computed.

\section{Experiments}\label{sec::exp}

\subsection{Data}
\label{sec:data}

The scope for these experiments regards English speaking adult datasets across three emotion types: one acted dataset, eNTERFACE \cite{enterface},  one natural dataset, MOSEI \cite{mosei}, and one elicited dataset, IEMOCAP \cite{iemocap}. An overview of the emotion classifications represented in each dataset are each described briefly below. For each dataset, the big-six emotions \cite{ekman1992argument} are considered in training and testing: \textit{happy}, \textit{sad}, \textit{anger}, \textit{surprise}, \textit{disgust} and \textit{fear}.

eNTERFACE (ENT) consists of roughly 1 hour of acted English utterances \cite{enterface}. The training set is comprised of 38 speakers and the testing set contains the remaining 5 speakers. The data is split by 8 female speakers and 35 male speakers from 14 different nations. 
%
%

IEMOCAP (IEM6) comprises of over 12 hours of US-English utterances from 10 speakers (5 female and 5 male) \cite{iemocap}. There are five dyadic sessions (between two speakers) which are specifically scripted or contrived to elicit certain emotions.
The training data consists of the first 4 sessions (4 speakers) and the last session is split for the test set (2 speakers). 
It is common for IEMOCAP to be evaluated as four classes: \textit{happy}, \textit{sad}, \textit{anger} and \textit{neutral} (where \textit{excitement} is combined with \textit{happy}).
This test set will be referred to as IEM4.
%

MOSEI (MOS) is the largest sentiment and emotion dataset with approximately 65 hours of data and more than 1000 speakers \cite{mosei}. Data is collected from YouTube and the videos are not specifically designed as an emotion dataset so the emotional speech is seen as natural. %
The official training, validation and test splits for the ACL 2018 conference have been considered, where the training and validation sets are combined for training. These can be found at \url{https://github.com/A2Zadeh/CMU-MultimodalSDK/blob/master/mmsdk/mmdatasdk/dataset/standard_datasets/CMU_MOSEI/cmu_mosei_std_folds.py}.

\subsection{Features}

Experiments from \cite{milner2019cross} showed how sequence-based SER systems performed best in terms of unweighted and weighted accuracy with 23-dimensional log-Mel filterbank features.

\subsection{Implementation}

The \textit{BLSTMATT} contains two hidden layers of 512 nodes each. The output layer (size 1024) is passed into the attention mechanism computing a context vector (size 128), which is projected to 1024 nodes. This is then fed into the emotion classifier which linearly projects to the 6 classes. The cross-entropy loss function is applied, which is preceded by a $softmax$ layer. 
The \textit{BLSTMATT} produces a variable length attention vector based on the input segment length, as mentioned in section \ref{ssec:blstmatt}. The attention vectors have been extracted and mapped with the phones and words in the input segments to be able to interpret the acoustic attention. 


\subsection{Evaluation} 

Unweighted accuracy (UA) and the weighted accuracy (WA) are the metrics typically applied for SER evaluation. The UA calculates accuracy in terms of the total correct predictions divided by total samples, which gives the same weight to each class:
\begin{equation}
    UA=\dfrac{TP+TN}{P+N}
\end{equation}
where $P$ is the number of correct positive instances (equivalent to $TP+FN$) and $N$ is the number of correct negative instances (equivalent to $TN+FP$).
As some of the datasets are imbalanced across the emotion classes, see Tables \ref{tab:my-table1}, \ref{tab:my-table2} and \ref{tab:my-table3}, the WA is calculated which weighs each class according to the number of samples in that class:
\begin{equation}
    WA=\dfrac{1}{2}(\dfrac{TP}{P}+\dfrac{TN}{N})
\end{equation}

Further details regarding the implementation of the scoring scripts can be found in \cite{beard2018multi}.




\subsection{Acoustic Context}

As discussed in Section \ref{sec:context} and \ref{sec:linguist}, the recognition of speech emotion is hypothesised to be influenced by overlapping perceptual acoustic cues consisting of variation in the phone boundaries. So, in theory, if the phone boundaries are shifted, the emotion classification may differ from the previous predicted emotion state that considered the whole segment. To further explore this hypothesis, the acoustic context is changed in the following series of experiments. 

Experiments are performed removing frames from the end and beginning of the original, whole test segments. In Tables \ref{tab:my-table1}, \ref{tab:my-table2} and \ref{tab:my-table3}, this is listed in the first column labelled `skip frames (left-right)' where a number of frames are skipped, or removed, from the left and right (left and right context) of each test segment. For example, 20-200 means 20 frames have been removed from the left context of each test segment and 200 frames have been removed from the right context of each test segment. Table \ref{tab:my-table1} shows the results where right frames are skipped, Table \ref{tab:my-table2} shows results where only left frames are skipped and Table \ref{tab:my-table3} shows results where both left and right frames are skipped.
If the length of a test utterance is less than the length of context frames, the test utterance remains unchanged. Therefore, when the skip context frames become longer, such as 200-100 (that means a total of 300 frames to be removed), only the test segments with more frames than 300 are used. The percentage of test corpora that is modified with the context is also reported. For example, in the SEGS\% column, 91.3\% means that 8.7\% of the test segments from the corresponding corpora remains the same due to shorter segment length and 91.3\% of the test segments are modified with the corresponding context. The weighted and unweighted accuracy are reported along with the change in the context length.

As the experiments consider context length variations, the baseline for this work is the result when no left or right context is removed. This is the first line in all Tables with context 0-0. It is the emotion modelling baseline where one emotion is given for each complete test utterance.
For further details about the validity of the \textit{BLSTMATT} model, please see work in \cite{milner2019cross} and \cite{jalal2020empirical}.


\begin{table*}[t!]
\caption{Cross-corpora results with variable context length, where right frames are skipped. }
\label{tab:my-table1}
\resizebox{\textwidth}{!}{%
\begin{tabular}{|l|cccc|cccc|cccc|}
\hline
\begin{tabular}[c]{@{}c@{}}Skip \\ Frames\\ (left-right)\end{tabular} &
  \multicolumn{4}{c|}{Unweighted Accuracy (UA \%)} &
  \multicolumn{4}{c|}{Weighted Accuracy (WA \%)} &
  \multicolumn{4}{c|}{\begin{tabular}[c]{@{}c@{}}Percentage of segments (SEGS \%) \\ with modified  context\end{tabular}} \\ \hline
Context &
  \multicolumn{1}{c|}{ENT} &
  \multicolumn{1}{c|}{IEM6} &
  \multicolumn{1}{c|}{IEM4} &
  MOS &
  \multicolumn{1}{c|}{ENT} &
  \multicolumn{1}{c|}{IEM6} &
  \multicolumn{1}{c|}{IEM4} &
  MOS &
  \multicolumn{1}{c|}{ENT} &
  \multicolumn{1}{c|}{IEM6} &
  \multicolumn{1}{c|}{IEM4} &
   MOS \\ \hline
  
0-0 &
  \multicolumn{1}{c|}{93.33} &
  \multicolumn{1}{c|}{69.06} &
  \multicolumn{1}{c|}{88.79} &
  73.30 &
  \multicolumn{1}{c|}{88.00} &
  \multicolumn{1}{c|}{64.57} &
  \multicolumn{1}{c|}{63.81} &
  54.29 &
  \multicolumn{1}{c|}{-} &
  \multicolumn{1}{c|}{-} &
  \multicolumn{1}{c|}{-} &
  - \\ \hline
0-30 &
  \multicolumn{1}{c|}{86.89} &
  \multicolumn{1}{c|}{68.73} &
  \multicolumn{1}{c|}{88.28} &
  73.55 &
  \multicolumn{1}{c|}{76.40} &
  \multicolumn{1}{c|}{63.82} &
  \multicolumn{1}{c|}{64.44} &
  54.76 &
  \multicolumn{1}{c|}{100.0} &
  \multicolumn{1}{c|}{100.0} &
  \multicolumn{1}{c|}{100.0} &
  100.0 \\ \hline
0-100 &
  \multicolumn{1}{c|}{82.22} &
  \multicolumn{1}{c|}{68.73} &
  \multicolumn{1}{c|}{87.94} &
  72.81 &
  \multicolumn{1}{c|}{68.00} &
  \multicolumn{1}{c|}{63.45} &
  \multicolumn{1}{c|}{61.21} &
  54.70 &
  \multicolumn{1}{c|}{91.3} &
  \multicolumn{1}{c|}{99.3} &
  \multicolumn{1}{c|}{99.5} &
  98.0 \\ \hline
0-200 &
  \multicolumn{1}{c|}{86.22} &
  \multicolumn{1}{c|}{68.09} &
  \multicolumn{1}{c|}{86.63} &
  71.53 &
  \multicolumn{1}{c|}{75.20} &
  \multicolumn{1}{c|}{62.08} &
  \multicolumn{1}{c|}{61.32} &
  53.63 &
  \multicolumn{1}{c|}{40.0} &
  \multicolumn{1}{c|}{77.1} &
  \multicolumn{1}{c|}{80.2} &
  89.4 \\ \hline
\end{tabular}%
}
\vspace{1.5mm}
\caption{Cross-corpora results with variable context length, where left frames are skipped. }
\label{tab:my-table2}
\resizebox{\textwidth}{!}{%
\begin{tabular}{|l|cccc|cccc|cccc|}
\hline
\begin{tabular}[c]{@{}c@{}}Skip \\ Frames\\ (left-right)\end{tabular} &
  \multicolumn{4}{c|}{Unweighted Accuracy (UA \%)} &
  \multicolumn{4}{c|}{Weighted Accuracy (WA \%)} &
  \multicolumn{4}{c|}{\begin{tabular}[c]{@{}c@{}}Percentage of segments (SEGS \%) \\ with modified  context\end{tabular}} \\ \hline
Context &
  \multicolumn{1}{c|}{ENT} &
  \multicolumn{1}{c|}{IEM6} &
  \multicolumn{1}{c|}{IEM4} &
  MOS &
  \multicolumn{1}{c|}{ENT} &
  \multicolumn{1}{c|}{IEM6} &
  \multicolumn{1}{c|}{IEM4} &
  MOS &
  \multicolumn{1}{c|}{ENT} &
  \multicolumn{1}{c|}{IEM6} &
  \multicolumn{1}{c|}{IEM4} &
  MOS \\ \hline
  
0-0 &
  \multicolumn{1}{c|}{93.33} &
  \multicolumn{1}{c|}{69.06} &
  \multicolumn{1}{c|}{88.79} &
  73.30 &
  \multicolumn{1}{c|}{88.00} &
  \multicolumn{1}{c|}{64.57} &
  \multicolumn{1}{c|}{63.81} &
  54.29 &
  \multicolumn{1}{c|}{-} &
  \multicolumn{1}{c|}{-} &
  \multicolumn{1}{c|}{-} &
  - \\ \hline
20-0 &
  \multicolumn{1}{c|}{91.56} &
  \multicolumn{1}{c|}{69.22} &
  \multicolumn{1}{c|}{88.74} &
  73.01 &
  \multicolumn{1}{c|}{84.80} &
  \multicolumn{1}{c|}{64.76} &
  \multicolumn{1}{c|}{63.83} &
  54.13 &
  \multicolumn{1}{c|}{100.0} &
  \multicolumn{1}{c|}{100.0} &
  \multicolumn{1}{c|}{100.0} &
  100.0 \\ \hline

30-0 &
  \multicolumn{1}{c|}{89.33} &
  \multicolumn{1}{c|}{69.66} &
  \multicolumn{1}{c|}{88.68} &
  72.97 &
  \multicolumn{1}{c|}{80.80} &
  \multicolumn{1}{c|}{65.03} &
  \multicolumn{1}{c|}{63.83} &
  54.18 &
  \multicolumn{1}{c|}{100.0} &
  \multicolumn{1}{c|}{100.0} &
  \multicolumn{1}{c|}{100.0} &
  100.0 \\ \hline
100-0 &
  \multicolumn{1}{c|}{84.44} &
  \multicolumn{1}{c|}{69.90} &
  \multicolumn{1}{c|}{87.66} &
  72.50 &
  \multicolumn{1}{c|}{72.00} &
  \multicolumn{1}{c|}{64.60} &
  \multicolumn{1}{c|}{61.00} &
  54.16 &
  \multicolumn{1}{c|}{91.3} &
  \multicolumn{1}{c|}{99.3} &
  \multicolumn{1}{c|}{99.5} &
  98.0 \\ \hline
200-0 &
  \multicolumn{1}{c|}{87.78} &
  \multicolumn{1}{c|}{69.38} &
  \multicolumn{1}{c|}{87.14} &
  71.98 &
  \multicolumn{1}{c|}{78.00} &
  \multicolumn{1}{c|}{63.47} &
  \multicolumn{1}{c|}{61.97} &
  54.07 &
  \multicolumn{1}{c|}{40.0} &
  \multicolumn{1}{c|}{77.1} &
  \multicolumn{1}{c|}{80.2} &
  89.4 \\ \hline
300-0 &
  \multicolumn{1}{c|}{92.89} &
  \multicolumn{1}{c|}{68.61} &
  \multicolumn{1}{c|}{86.86} &
  71.11 &
  \multicolumn{1}{c|}{87.20} &
  \multicolumn{1}{c|}{62.89} &
  \multicolumn{1}{c|}{61.66} &
  53.45 &
  \multicolumn{1}{c|}{8.7} &
  \multicolumn{1}{c|}{54.5} &
  \multicolumn{1}{c|}{57.8} &
  80.2 \\ \hline
\end{tabular}%
}

\vspace{1.5mm}
\caption{Cross-corpora results with variable context length, where left and right frames are skipped. }
\label{tab:my-table3}
\resizebox{\textwidth}{!}{%
\begin{tabular}{|l|cccc|cccc|cccc|}
\hline
\begin{tabular}[c]{@{}c@{}}Skip \\ Frames\\ (left-right)\end{tabular} &
  \multicolumn{4}{c|}{Unweighted Accuracy (UA \%)} &
  \multicolumn{4}{c|}{Weighted Accuracy (WA \%)} &
  \multicolumn{4}{c|}{\begin{tabular}[c]{@{}c@{}}Percentage of segments (SEGS \%) \\ with modified  context\end{tabular}} \\ \hline
Context &
  \multicolumn{1}{c|}{ENT} &
  \multicolumn{1}{c|}{IEM6} &
  \multicolumn{1}{c|}{IEM4} &
  MOS &
  \multicolumn{1}{c|}{ENT} &
  \multicolumn{1}{c|}{IEM6} &
  \multicolumn{1}{c|}{IEM4} &
  MOS &
  \multicolumn{1}{c|}{ENT} &
  \multicolumn{1}{c|}{IEM6} &
  \multicolumn{1}{c|}{IEM4} &
   MOS \\ \hline
  
0-0 &
  \multicolumn{1}{c|}{93.33} &
  \multicolumn{1}{c|}{69.06} &
  \multicolumn{1}{c|}{88.79} &
  73.30 &
  \multicolumn{1}{c|}{88.00} &
  \multicolumn{1}{c|}{64.57} &
  \multicolumn{1}{c|}{63.81} &
  54.29 &
  \multicolumn{1}{c|}{-} &
  \multicolumn{1}{c|}{-} &
  \multicolumn{1}{c|}{-} &
  - \\ \hline
20-200 &
  \multicolumn{1}{c|}{86.89} &
  \multicolumn{1}{c|}{68.73} &
  \multicolumn{1}{c|}{87.49} &
  71.39 &
  \multicolumn{1}{c|}{76.40} &
  \multicolumn{1}{c|}{63.11} &
  \multicolumn{1}{c|}{62.44} &
  53.81 &
  \multicolumn{1}{c|}{38.7} &
  \multicolumn{1}{c|}{71.6} &
  \multicolumn{1}{c|}{73.4} &
  87.6 \\ \hline
200-100 &
  \multicolumn{1}{c|}{92.67} &
  \multicolumn{1}{c|}{68.25} &
  \multicolumn{1}{c|}{86.58} &
  71.30 &
  \multicolumn{1}{c|}{86.80} &
  \multicolumn{1}{c|}{61.83} &
  \multicolumn{1}{c|}{61.42} &
  54.00 &
  \multicolumn{1}{c|}{8.7} &
  \multicolumn{1}{c|}{54.5} &
  \multicolumn{1}{c|}{57.8} &
  80.2 \\ \hline

\end{tabular}%
}

\end{table*}

\begin{figure*}[t!]
	 
		\centering
		\includegraphics[width=0.9\linewidth,trim={1cm 0.5cm 1cm 1.75cm}, clip]{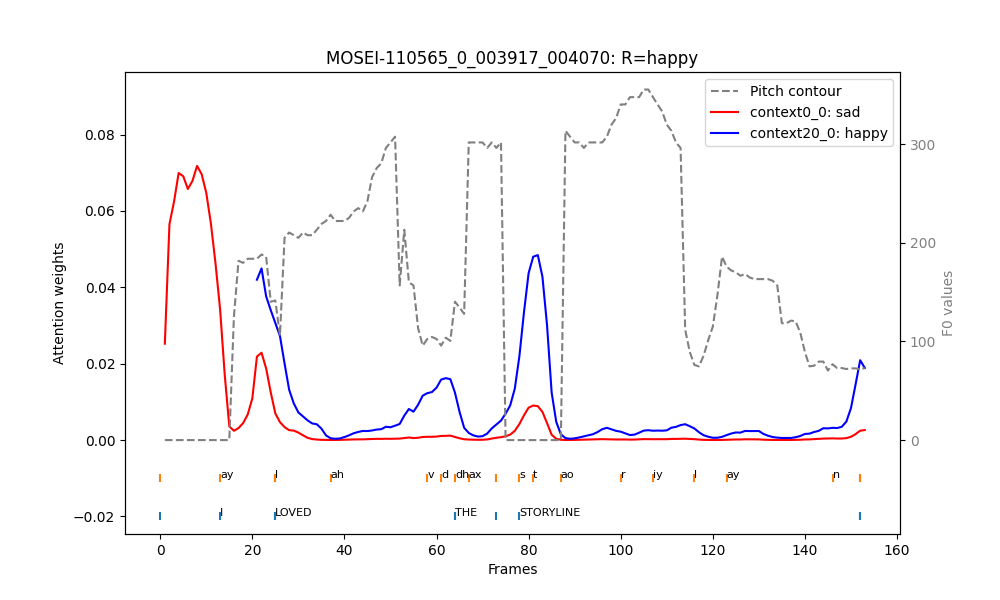}
		\caption{A \textit{happy} MOS utterance with no context removed mislabelled as \textit{sad} compared to 20 left frames removed correctly labelled as \textit{happy}, along with the pitch contour 		}
        \label{fig:attplots1}
\end{figure*}

\begin{figure*}[t!]
	    \centering
		\includegraphics[width=0.9\linewidth,trim={1cm 0.5cm 1cm 1.75cm}, clip]{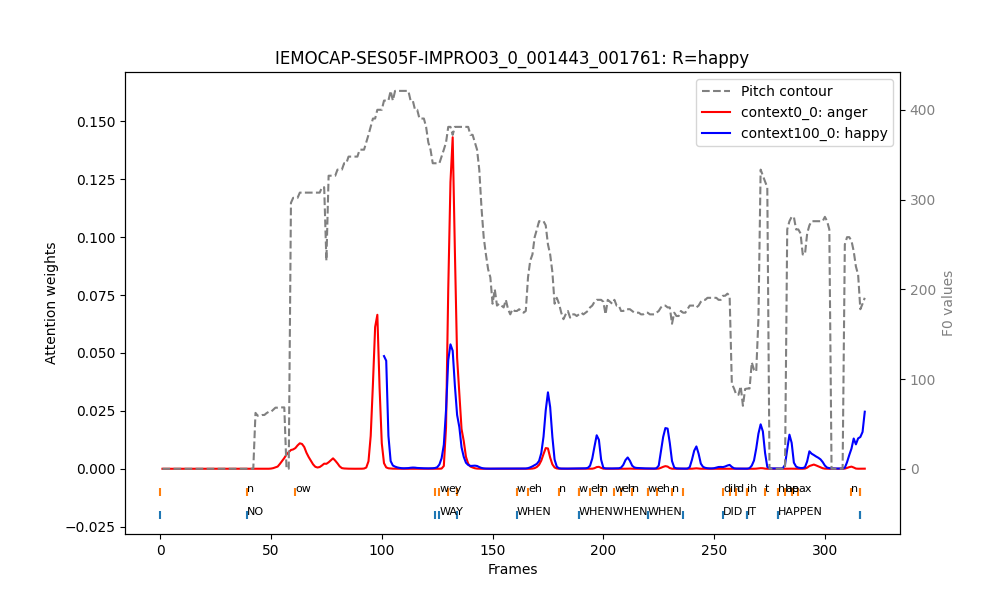}
		\caption{A \textit{happy} IEM4 utterance with no context removed mislabelled as \textit{anger} compared to 100 right frames removed correctly labelled as \textit{happy}, along with the pitch contour} 
        \label{fig:attplots2} 
\end{figure*}

\section{Results}
\label{sec:result}
The experimental results in Tables \ref{tab:my-table1}, \ref{tab:my-table2} and \ref{tab:my-table3} suggest that the SER results would change when either the left, right or both contexts are changed. For example, the model tested on MOS has a UA of 73.3\% without changing the context length but upon skipping the context right 100 frames, skip frames 0-100 in Table \ref{tab:my-table1}, the UA degrades. The same observation of UA degradation occurs when skipping left frames or both left and right frames, while there is a slight improvement in UA when skipping 30 right frames. In the case of skip frames 0-30, removing 30 ms from the end of the segment modifies 100\% of the segments across all the testsets. The results for ENT and IEM4 are worse for both UA and WA, but for MOS the performance improves. For IEM6, the UA degrades whereas the WA improves. The majority of the results across all the datasets degrade upon varying the context length due to the target label, supplied with those segments, being a fixed discrete emotion category. This finding corroborates the initial hypotheses that speech emotion is not a fixed entity that remains the same over the whole audio segment, and that it is subject to be distributed over different overlapping shorter context queues. 

To observe the relationship between the SER results and the hypotheses regarding the acoustic segments in more detail, the attention weights were extracted for each test utterance and mapped to the aligned words and phones. Additionally, the pitch contour was calculated to understand the pitch correlation with respect to the prosodic utterance using the algorithm found at \url{https://github.com/google/REAPER}. The attention maps for a sample of the test utterances are presented in Figures \ref{fig:attplots1} and \ref{fig:attplots2}: the former from the MOSEI corpus and the latter from the IEMOCAP corpus. Figure \ref{fig:attplots1} shows that the attention projection drifts while changing the phone boundaries from the same audio segment and therefore the emotion state also changes. With context 0-0, the model incorrectly predicts the emotion \textit{sad} (attention weights in Figure \ref{fig:attplots1} indicated by red line) whereas removing 20 left frames helps the model correctly predict the emotion \textit{happy} (attention weights in Figure \ref{fig:attplots2} indicated by blue line). The attention weights focus more strongly on different portions of the test utterance. Similar behaviour can be seen in Figure \ref{fig:attplots2} where skipping 100 left frames allows the model to make the correct prediction. 




\section{Discussion and Conclusions}\label{sec:disscussion}
\label{sec:discussions}



In some datasets, such as ENT and IEM4, the SER change is not very prominent, as can be seen from the results listed for WA and UA. This is potentially due to several reasons. The first is that some segments in particular datasets have a length shorter than 200 frames, and these segments remained unchanged during the context modification. So to properly interpret the results, the percentage of data that is modified with context at each experiment should be taken into account. Secondly, the acoustic \textit{BLSTMATT} model perception is a direct result of the relationship between the training data and the corresponding labels given by the annotators, which could have added bias factors and add to recognition uncertainty. To attempt to mitigate the inherent biases and to attempt to generalise the model perception cues for these experiments, the model is trained with four different corpora consisting of acted, natural and elicited emotions. Consequently, it is argued that the results corroborate the argument that an continuous approach to emotion recognition is the optimal strategy based on observed acoustic stimuli shift. This work is an attempt to bridge the gap that current SER models have, by explaining the SER model's internal intricacies and how the representations correspond with acoustic segments.

Figures \ref{fig:attplots1} and \ref{fig:attplots2} show the attention weight propensity towards the vowel based regions. This corroborates with the claims from the linguistic and cognitive theories about speech emotion recognition and CV boundaries, as proposed in Section \ref{sec:linguist}, that consonants play a decisive role in word meaning but vowels are more responsible for the emotion perception cues as a result of harmonic variations and stressed regions. The vowels are observed to change the CV boundaries and the context cues for emotion perception causing many hard boundaries to be redundant. This suggests the cues for phone boundaries and acoustic context can share information relative to the perceived emotion state. 

For the IEM6 dataset, when the context lengths were skipped left frames, there was a slight improvement in UA or WA, while recognition with the MOS and IEM4 datasets improved slightly when context lengths were skipped right frames. These results highlight that where context cues vary in length, it is possible for the acoustic segments to contain more than one distinct emotion state. As the UA and WA vary positively and negatively according to context lengths, this suggests overlapping regions where the acoustic stimuli are more or less informative regarding the emotion state. As future speech emotion datasets are compiled and annotated, if the labels for emotion classes were adjusted to allow for overlapping categories, this could potentially aid the recognition performance of current and future developed models. These results and insights can also be used to modify computational models and mechanisms that are able to adapt and recognise emotion from various speech domains to be more in-line with the psycholinguistic theories.


In the current trend of SER models, emotion labels are treated as discrete labels attributed over a whole segment. The problem explored in this research, suggests that this approach assumes that an utterance's global attributes correlate with the local characteristics over different time frames in the same segment for learning one discrete emotion category. This is observed to not be the case most of the time. Vowel-consonant envelopes rapidly change over time, attributing to different acoustic context. Hence the paralinguistic cue also changes with acoustic context. The results listed in Section \ref{sec:result} demonstrate this argument. Moreover, by treating acoustic segments and emotion correspondence as a context-oriented continuous relationship, this should aid emotion recognition models across languages and dialects due to the distribution of acoustic boundaries across models trained on various emotion data. As a result, it could be possible to learn the variability of acoustic context in speech emotions rather than the variability of acoustic segments in speech emotions. 

Future development of this framework will enable improved emotion modelling by understanding the intermediate representations and relating audio data with the computational models. Furthermore, it will help create more accurate annotations for emotion labels, improving SER corpora generation. 

This work argues that discrete categorical emotion classification should not be the preferred approach to develop future SER models as it has been observed that emotion cues present as a distributed event, corroborating directly with cognitive linguistic theory that it is also continuous to recognise. Finding a suitable approach for accurate modelling of emotion states should be the aim of future research.

\bibliographystyle{IEEEtran}
\bibliography{main}

\end{document}